\documentclass[pdftex,twocolumn,epjc3]{svjour3}          

\RequirePackage[T1]{fontenc}

\smartqed  

\RequirePackage{graphicx}
\RequirePackage{mathptmx}      
\RequirePackage{flushend}
\RequirePackage[numbers,sort&compress]{natbib}
\RequirePackage[colorlinks,citecolor=blue,urlcolor=blue,linkcolor=blue]{hyperref}

\usepackage{amsmath,amssymb,graphicx}

\journalname{TTMP}

	[2011/08/26 v1.2
	style option for the EPJC journal]
	
	\AtEndOfClass{\@input{svepjc3.clo}}
	\gdef\@epjcloaded{}

	\AtBeginDocument{%
	\def\@xhline{\ifx\reserved@a\hline
		\vskip\doublerulesep\vskip-\arrayrulewidth\fi
		\ifnum0=`{\fi}%
		\noalign{\vskip2\p@}}%
}

\begin{document}
	
	\title{Linear independence of field equations in the Brans-Dicke theory
 }

	\author{E. Ahmadi-Azar\thanksref{e1}
		\and
        K. Atazadeh\thanksref{e2}
        \and
		A. Eghbali\thanksref{e3} 
	}

	\thankstext{e1}{e.ahmadi.azar@azaruniv.ac.ir}
	\thankstext{e2}{atazadeh@azaruniv.ac.ir}
    \thankstext{e3}{eghbali978@gmail.com}
	
	\institute{Department of Physics, Azarbaijan Shahid Madani University, Tabriz, Iran}
	
	\date{Received: date / Accepted: date}

	\maketitle

\section*{Abstract}
In solving the Brans-Dicke (BD) equations in the BD theory of gravity, their linear independence is important.
This is due to fact that in solving these equations in cosmology, if the number of unknown quantities is equal to the number of independent equations,
then the unknowns can be uniquely determined.
In the BD theory, the tensor field $g_{\mu \nu}$ and the BD scalar field $\varphi$
are not two separate fields, but they are coupled together.
The reason behind this is a corollary that proposed by V. B. Johri and D. Kalyani in cosmology,
which states that the cosmic scale factor of the universe, $a$, and the BD scalar field $\varphi$
are related by a power law. Therefore, when the principle of least action is used to derive the BD equations, the variations $\delta g^{\mu \nu}$
and $\delta \varphi$ should not be considered as two independent dynamical variables. So, there is a
constraint on $\delta g^{\mu \nu}$ and $\delta \varphi$ that causes the number of independent BD equations to decrease by one unit,
in such a way that in the equations that have been known as BD equations, one of them is redundant.
In this paper, we prove this issue, that is, we show that one of these equations,
which we choose as the modified Klein-Gordon equation, is not an independent equation,
but a result establishing other BD equations, the law of conservation of energy-momentum of matter and Bianchi's identity.
Therefore, we should not look at the modified Klein-Gordon equation as an independent field equation in the BD theory,
but rather it is included in the other BD equations and should not be mentioned separately as one of the BD equations once again.
\\\\
{\bf Keywords:}  Brans-Dicke equations; Modified Klein-Gordon equation; Bianchi identity
\section{\label{I} Introduction}

In solving the BD equations in the BD theory  \cite{r1,r2}, their linear independence is important.
This is due to the fact that the number of independent equations must be equal to the number of unknowns in their solutions.
Actually, in BD theory, BD equations form a system of coupled nonlinear second-order differential equations.
One of these equations is the modified Klein-Gordon equation and the other equations are generalized Einstein's field equations (EFEs).
In the spatially flat $(k=0)$ FLRW cosmological model \cite{r3,r4,r5,r6,r7,r8,r9} for a universe that is a perfect fluid with the equation of state
\begin{eqnarray}
p=w \rho,\label{eq1}
\end{eqnarray}
where  $-1\leq w \leq 1$ , what we seek from solving the BD equations is that to find four quantities:
the cosmic scale factor $a$, the BD scalar field $\varphi$, the energy density of the universe $\rho$ and its pressure $p$.
To determine these quantities as functions of cosmic time $t$, we must have four differential-algebraic equations.
Then we can uniquely determine the unknown quantities $a$, $\varphi$, $\rho$ and $p$ as functions of the cosmic time $t$
from solutions of this system of differential-algebraic equations.

The Johri-Kalyani's corollary requires that quantities $a$ and $\varphi$ are not independent of each other, but they are related by a power law as \cite{r6,r7,r8,r9,r10}
\begin{eqnarray}
\varphi a^n ={\cal C},\label{eq2}
\end{eqnarray}
where $n$ is an adjustable parameter and ${\cal C}$ is a constant.
This law first introduced into cosmology by Dehnen and Obreg\'{o}n \cite{rr11} as a hypothesis to solve the BD equations.
Later, following Dehnen and Obreg\'{o}n, authors \cite{r7,r8,r9,rr12,rr13} used this assumption to solve the BD equations.
Finally, in 1994 Johri and Kalyani \cite{r6} proved that this relation should not be seen as an assumption but rather a result of the constancy of the deceleration parameter of the universe.
Accordingly, in this article we have called this power law as Johri-Kalyani's corollary.
Thus, in the FLRW cosmological model, in addition to the equation of state (\ref{eq1}), we have also the power law (\ref{eq2})
as an algebraic equation between the unknowns of the problem.
So, to determine the unknowns of the problem, we only need two differential equations that together with these two algebraic
equations form a system with four equations
for four unknowns $a$, $\varphi$, $\rho$ and $p$.
Based on this, the BD equations must contain two coupled independent second-order differential equations.

As a result, in order to solve the FLRW cosmological model,
the BD equations must be three differential equations. One of them must depend on the other two equations.
In other words, from the set of BD equations, one of them is repeated. Our goal in this paper is that to show in the BD theory, one of these equations,
which we choose here, the modified Klein-Gordon equation \cite{r11,r12}, is not an independent equation, but a result of the establishment of other BD equations.
For this purpose, we first derive the BD equations by using the principle of least action for the BD-action.
\section{\label{II} BD theory}

The BD-action in Jordan's frame is given by \cite{r11,r12,r13,r14,r15,r16,r17}
\begin{eqnarray}\label{eq3}
I_{_{BD}}&=&\frac{1}{16\pi}\int_{\cal M} ~d^4x \sqrt{-g} \Big(\varphi {\cal R}-\frac{\omega}{\varphi}~g^{\mu \nu}~\nabla_{\mu} \varphi~\nabla_{\nu} \varphi\nonumber\\
~~~~~~~~~~&& -V(\varphi) +16\pi~{\mathcal{L}}_{_{M}}\Big),
\end{eqnarray}
where $\varphi$ is the BD scalar field, $\omega$ is the adjustable BD coupling parameter, $V(\varphi)$ is the potential energy of the field
$\varphi$ and ${\cal R}$ is the Ricci scalar of the space-time manifold ${\cal M}$ with the local coordinates
$x^\mu = (x^0, x^1, x^2, x^3)$. Also,  $g =\det g_{\mu \nu}$ and $\nabla_\mu$ denotes the covariant derivative operator in the space-time
and finally, ${\mathcal{L}}_{_{M}} := {\mathcal{L}}_{_{M}} (g_{\mu \nu} , \partial_\rho g_{\mu \nu})$ is the Lagrangian density of the matter which is minimally coupled to the BD scalar field $\varphi$.
Moreover, the BD scalar field $\varphi$ inversely proportional to the effective gravitational constant $G_{eff}$  by the relation
\cite{r7,r8,r12,r16,r17}
\begin{eqnarray}\label{eq4}
\varphi =\frac{1}{G_{eff}} \frac{4+2 \omega}{3+2 \omega},
\end{eqnarray}
where ${G_{eff}}$ is equal to the Newton's gravitational constant $G = 6.67 \times 10^{-8}  ~cm^3~ g^{-1} ~s^{-2}$  in the limit when $\omega$ tends to infinity.

It should be noted that Eq. (\ref{eq3}) is the original action of the BD theory. 
In 2011, the generalization of this action was presented by S. Nojiri and S. D. Odintsov in \cite{r21,r22} as follows:
\begin{eqnarray}\label{eq5}
I_{_{NO}}&=&\frac{1}{16\pi}\int_{\cal M} ~d^4x \sqrt{-g} \Big(e^{\alpha(\varphi)} {\cal R}-\frac{\omega(\varphi)}{\varphi}~g^{\mu \nu}~\nabla_{\mu} \varphi~\nabla_{\nu} \varphi\nonumber\\
~~~~~~~~~~&& -V(\varphi) +16\pi~{\mathcal{L}}_{_{M}}\Big),
\end{eqnarray}
where $e^{\alpha(\varphi)}$  and $\omega(\varphi)$ are some appropriate functions of the BD scalar field $\varphi$. 
The above action, Eq. (\ref{eq5}), was applied to dark energy problem \cite{r23}. But in the present paper we focus on the original BD action, Eq. (\ref{eq3}).

Similar to the theory of general relativity (GR), the definition of energy-momentum tensor (EMT) of the matter
\begin{eqnarray}\label{eq5.1}
T_M^{\mu \nu} =\frac{2}{\sqrt{-g}} \frac{\partial}{\partial g_{\mu \nu}} \Big(\sqrt{-g} {\mathcal{L}}_{_{M}}\Big),
\end{eqnarray}
and its conservation
\begin{eqnarray}\label{eq6}
\nabla_{\mu} T_M^{\mu \nu} =0,
\end{eqnarray}
also hold in the BD theory \cite{r7,r8,r9,r11,r12,r16,r18}.

By varying the BD action, Eq. (\ref{eq3}), with respect to the dynamical variables,
BD scalar field $\varphi$ and metric tensor $g^{\mu \nu}$, and by using the least action principle for the BD action, i.e. $\delta I_{BD}=0$,
and this fact that the variations of $\delta g^{\mu \nu}$ and $\delta \varphi$ are arbitrary,
then we get the following equations, respectively \cite{r9,r11,r12,r17}
\begin{eqnarray}
&&\frac{2 \omega}{\varphi } \square {\varphi} + {\cal R} -\frac{\omega}{\varphi^2 } \nabla^{\mu} \varphi \nabla_{\mu} \varphi-\frac{dV}{d \varphi}=0,\label{eq7}\\
&&{\cal R}_{\mu \nu}-\frac{1}{2} g_{\mu \nu} {\cal R} =\frac{8 \pi}{\varphi}{T_M}_{\mu \nu} +\frac{1}{\varphi} \big(\nabla_{\mu} \nabla_{\nu} \varphi
-g_{\mu \nu}  \square {\varphi}\big)\nonumber\\
&&~~~~~~~~~~~~+\frac{\omega}{\varphi^2 }\big(\nabla_{\mu} \varphi \nabla_{\nu} \varphi
-\frac{1}{2} g_{\mu \nu} \nabla^{\rho} \varphi \nabla_{\rho} \varphi\big) -\frac{V}{2 \varphi} g_{\mu \nu},\label{eq8}
\end{eqnarray}
where ${\cal R}_{\mu \nu}$ is the Ricci tensor and $\square $ is the covariant d'Alemberian
operator of the metric tensor $g_{\mu \nu}$, which is defined by
\begin{eqnarray}\label{eq9}
 \square :=\nabla^{\rho}  \nabla_{\rho} =\frac{1}{\sqrt{-g}} \partial_{\mu}\big({\sqrt{-g}} g^{\mu \nu}  \partial_{\nu}\big).
\end{eqnarray}
We note that Eq. (\ref{eq8}) is a generalization of the Einstein's field equations ${\cal R}_{\mu \nu}-\frac{1}{2} g_{\mu \nu} {\cal R} + \Lambda g_{\mu \nu}=8 \pi G {T_M}_{_{\mu \nu}}$.
By performing contraction on Eq. (\ref{eq8}), we obtain
\begin{eqnarray}\label{eq10}
{\cal R} = -\frac{8 \pi {T_M}^{\hspace{-1mm}\lambda}_{\lambda}}{\varphi} +\frac{\omega}{\varphi^2 } \nabla^{\mu} \varphi \nabla_{\mu} \varphi +\frac{3 \square {\varphi}}{\varphi} + \frac{2V}{\varphi},
\end{eqnarray}
where ${T_M}^{\hspace{-1mm}\lambda}_{_\lambda}:= g^{\mu \nu} {T_M}_{_{\mu \nu}}$ is the trace of the energy-momentum tensor of the ordinary matter fields.
By substituting Eq. (\ref{eq10}) into Eq. (\ref{eq7}) one gets
\begin{eqnarray}\label{eq11}
\square {\varphi} = \frac{1}{3+2 \omega} \big(8\pi {T_M}^{\hspace{-1mm}\lambda}_{_\lambda} + {\varphi} \frac{dV}{d \varphi} -2 V\big).
\end{eqnarray}
This equation together with Eq. (\ref{eq8}) form a system of coupled second-order non-linear differential equations which are called the general form of the BD equations in the BD theory of gravity.
Notice that Eq.  (\ref{eq11}) is known as the modified Klein-Gordon equation.
Furthermore,  Eq. (\ref{eq8}) is sometimes called BD field equations.

\section{\label{III} Linear independence of the field equations in the BD theory}

The purpose of this section is to show that in the BD theory, the Klein-Gordon equation, Eq. (\ref{eq11})
is not an independent equation; but it can be derived from the BD field equations, Eq. (\ref{eq8}).
For this purpose, we take the covariant derivative from both sides of Eq. (\ref{eq8}) with respect to the general coordinate $x^\nu$, giving us
\begin{eqnarray*}
\nabla_\nu \Big({\cal R}^{\mu \nu}-\frac{1}{2} g^{\mu \nu} {\cal R}\Big) = \nabla_\nu \Big[\frac{8 \pi}{\varphi}T_M^{\mu \nu } +\frac{1}{\varphi} \big(\nabla^{\mu} \nabla^{\nu} \varphi-g^{\mu \nu}  \square {\varphi}\big)
\end{eqnarray*}
\vspace{-4mm}
\begin{eqnarray}
~~~~~~~~~~+\frac{\omega}{\varphi^2 }\big(\nabla^{\mu} \varphi \nabla^{\nu} \varphi-\frac{1}{2} g^{\mu \nu} \nabla^{\rho} \varphi \nabla_{\rho} \varphi\big) -\frac{V}{2 \varphi} g^{\mu \nu}\Big].\label{eq12}
\end{eqnarray}
Then, using the Bianchi's identity
$\nabla_\nu G^{\mu \nu}=0$, ($G_{\mu \nu}= {\cal R}_{\mu \nu}-\frac{1}{2} {\cal R} g_{\mu \nu}$ is Einstein's tensor),
the left side of Eq. (\ref{eq12}) becomes zero.
On the right side, according to the EMT conservation law of the matter, Eq. (\ref{eq6}), the divergence of the EMT is zero.
Finally, we obtain
\begin{eqnarray}
&&-\frac{8 \pi } {\varphi^2} T_M^{\mu \nu } \nabla_\nu {\varphi} -\frac{2 \omega} {\varphi^3}\Big(\nabla^\mu {\varphi} \nabla^\nu {\varphi}
-\frac{1}{2} g^{\mu \nu} \nabla^\rho {\varphi} \nabla_\rho {\varphi}\Big) \nabla_\nu {\varphi}~~~~\nonumber\\
&& + \frac{\omega} {\varphi^2}\Big[(\nabla_\nu \nabla^\mu {\varphi}) \nabla^\nu {\varphi} + (\nabla_\nu  \nabla^\nu {\varphi})\nabla^\mu   {\varphi}
- \frac{1}{2}g^{\mu \nu} (\nabla_\nu \nabla^\rho {\varphi}) \nabla_\rho {\varphi} \nonumber\\
&&  - \frac{1}{2}g^{\mu \nu} \nabla^\rho {\varphi} (\nabla_\nu \nabla_\rho {\varphi})\Big] - \frac{1} {\varphi^2} (\nabla^\mu \nabla^\nu {\varphi} -g^{\mu \nu}  \square {\varphi}) \nabla_\nu {\varphi} \nonumber\\
&&  + \frac{1} {\varphi} \Big[\nabla_\nu (\nabla^\mu \nabla^\nu {\varphi}) -g^{\mu \nu} \nabla_\nu (\square {\varphi})\Big] \nonumber\\
&&+ (\frac{V}{2 \varphi^2}
- \frac{1}{2 \varphi} \frac{d V}{d \varphi})g^{\mu \nu} \nabla_\nu {\varphi}=0.\label{eq13}
\end{eqnarray}
By doing some tensor calculations,  Eq. (\ref{eq13}) can be simplified as follows:
\begin{eqnarray}
&&-\frac{8 \pi } {\varphi^2} T_M^{\mu \nu } \nabla_\nu {\varphi}
-\frac{\omega} {\varphi^3} (\nabla^\rho {\varphi} \nabla_\rho {\varphi}) \nabla^\mu {\varphi}\nonumber\\
&&+ \frac{\omega} {\varphi^2} (\square {\varphi}) \nabla^\mu {\varphi}
-\frac{1} {\varphi^2}  (\nabla^\mu \nabla^\nu {\varphi}) \nabla_\nu {\varphi}\nonumber\\
&&+ \frac{1} {\varphi^2} (\square {\varphi}) \nabla^\mu {\varphi} + \frac{1} {\varphi} \Big[\square (\nabla^\mu {\varphi})
-\nabla^\mu  (\square {\varphi})\Big]\nonumber\\
&&+ (\frac{V}{2 \varphi^2}
- \frac{1}{2 \varphi} \frac{d V}{d \varphi}) \nabla^\mu {\varphi}=0.\label{eq14}
\end{eqnarray}
In order to calculate expression $\square (\nabla^\mu {\varphi})
-\nabla^\mu  (\square {\varphi})$ in the above equation, let us consider the following tensor identity \cite{r15,r18}
\begin{eqnarray}
  \nabla_\mu  \nabla_\nu V^\lambda -    \nabla_\nu  \nabla_\mu V^\lambda = V^\sigma {\cal R}^\lambda_{~~\sigma \mu \nu}, \label{eq15}
\end{eqnarray}
where $V^\lambda$ is an arbitrary four-vector and $ {\cal R}^\lambda_{~~\sigma \mu \nu}$ is the Riemann curvature tensor.
By putting $\lambda = \nu$ in the above identity, we then get
\begin{eqnarray}
\nabla_\mu  \nabla_\nu V^\nu-    \nabla_\nu  \nabla_\mu V^\nu &=& V^\sigma {\cal R}^\nu_{~~\sigma \mu \nu}, \nonumber\\
&=& -V^\sigma {\cal R}^\nu_{~~\sigma \nu \mu} \nonumber\\
&=& -V^\sigma  {\cal R}_{\sigma \mu} \nonumber\\
&=&-V_\sigma  {\cal R}^{\sigma}_{~\mu},\label{eq16}
\end{eqnarray}
where ${\cal R}_{\mu \nu} := {\cal R}^\lambda_{~~\mu \lambda \nu}$ is the Ricci tensor. In Eq. (\ref{eq16}), we define co-vector
$V_\sigma = \nabla_\sigma \varphi$, where $\varphi$ is the BD scalar field. Accordingly, we get the following relation
\begin{eqnarray}
\nabla_\mu  (\nabla_\nu \nabla^\nu  \varphi) -\nabla_\nu  (\nabla_\mu \nabla^\nu  \varphi) = -\nabla_\sigma \varphi {\cal R}^{\sigma}_{~\mu},
\label{eq17}
\end{eqnarray}
which can be written as follows:
\begin{eqnarray}
\nabla_\nu (\square {\varphi}) -\square (\nabla_\nu {\varphi}) = -(\nabla_\mu \varphi) {\cal R}^{\mu}_{~\nu}.
\label{eq18}
\end{eqnarray}
For the next use, it is useful to write the above equation in the following form
\begin{eqnarray}
\nabla^\nu (\square {\varphi}) -\square (\nabla^\nu {\varphi}) = -(\nabla_\mu \varphi) {\cal R}^{\mu \nu}.
\label{eq19}
\end{eqnarray}
From the contraction of Eq. (\ref{eq8}) one can obtain the Ricci scalar $\cal R$, giving us
\begin{eqnarray}
&&{\cal R}^{\mu}_{~\nu}-\frac{1}{2} {\delta}^{\mu}_{~\nu} {\cal R} =\frac{8 \pi}{\varphi}{T_M^\mu}_{\nu} +\frac{1}{\varphi} \big(\nabla^{\mu} \nabla_{\nu} \varphi -{\delta}^{\mu}_{~\nu}   \square {\varphi}\big)\nonumber\\
&&~~~~~~~~~~~~+\frac{\omega}{\varphi^2 }\big(\nabla^{\mu} \varphi \nabla_{\nu} \varphi
-\frac{1}{2} {\delta}^{\mu}_{~\nu}  \nabla^{\rho} \varphi \nabla_{\rho} \varphi\big) -\frac{V}{2 \varphi}{\delta}^{\mu}_{~\nu}.\label{eq20}
\end{eqnarray}
By putting $\mu = \nu$ in the above equation and then by using the fact that ${\cal R}^{\mu}_{~\mu} = {\cal R}$ and
${\delta}^{\mu}_{~\mu} = 4$ we arrive at Eq. (\ref{eq10}).

Now, in Eq. (\ref{eq8}), instead of the Ricci scalar ${\cal R}$, we put its value from Eq. (\ref{eq10})
and then calculate the tensor ${\cal R}^{\mu \nu}$ from the resulting equation, giving us
\begin{eqnarray}
&&{\cal R}^{\mu \nu}-\frac{1}{2} {g}^{\mu \nu} \Big(- \frac{8 \pi}{\varphi}{T_M^\lambda}_{\lambda} +\frac{3}{\varphi} \square {\varphi}
+\frac{\omega}{\varphi^2} \nabla^{\lambda} \varphi \nabla_{\lambda} \varphi+\frac{2 V}{\varphi}\Big) \nonumber\\
&&~~~~~~~~~~~~=\frac{8 \pi}{\varphi}{T_M^{\mu \nu}} +\frac{\omega}{\varphi^2 }\big(\nabla^{\mu} \varphi \nabla^{\nu} \varphi
-\frac{1}{2} {g}^{\mu \nu}  \nabla^{\rho} \varphi \nabla_{\rho} \varphi\big)\nonumber\\
&&~~~~~~~~~~~~+\frac{1}{\varphi} \big(\nabla^{\mu} \nabla^{\nu} \varphi -{g}^{\mu \nu}  \square {\varphi}\big) -\frac{V}{2 \varphi}{g}^{\mu \nu}.\label{eq22}
\end{eqnarray}
After some simplification, we then obtain
\begin{eqnarray}
&&{\cal R}^{\mu \nu} = - \frac{4 \pi}{\varphi} {T_M^\lambda}_{\lambda}{g}^{\mu \nu} +\frac{8 \pi}{\varphi}{T_M^{\mu \nu}}
   +\frac{\omega}{\varphi^2 } \nabla^{\mu} \varphi \nabla^{\nu} \varphi \nonumber\\
 &&~~~~~~~~~~~~~~~~~~~~+\frac{1}{2 \varphi} \square {\varphi} {g}^{\mu \nu} + \frac{V}{2 \varphi}{g}^{\mu \nu}
+ \frac{1}{\varphi} \nabla^{\mu}  \nabla^{\nu} \varphi. \label{eq23}
\end{eqnarray}
By putting the tensor ${\cal R}^{\mu \nu}$ from the above equation into Eq. (\ref{eq19}), we get the following equation
\begin{eqnarray}
&&\nabla^\mu (\square {\varphi}) -\square (\nabla^\mu {\varphi}) ~~~~~~~~\nonumber\\
&&~~~~~~~~~~~~~= - \Big[- \frac{4 \pi}{\varphi} {T_M^\lambda}_{\lambda}{g}^{\mu \nu}
+\frac{8 \pi}{\varphi}{T_M^{\mu \nu}} \nonumber\\
&&~~~~~~~~~~~~~~+\frac{\omega}{\varphi^2 } \nabla^{\mu} \varphi \nabla^{\nu} \varphi +\frac{1}{2 \varphi} \square {\varphi} {g}^{\mu \nu}\nonumber\\
&&~~~~~~~~~~~~~~+ \frac{V}{2 \varphi}{g}^{\mu \nu} + \frac{1}{\varphi} \nabla^{\mu}  \nabla^{\nu} \varphi\Big] \nabla_\nu \varphi.
\label{eq24}
\end{eqnarray}
From the combination of Eqs. (\ref{eq14}) and (\ref{eq24}), one may get the following equation
\begin{eqnarray}
&&-\frac{8 \pi } {\varphi^2} T_M^{\mu \nu } \nabla_\nu {\varphi}
-\frac{\omega} {\varphi^3} (\nabla^\rho {\varphi} \nabla_\rho {\varphi}) \nabla^\mu {\varphi}\nonumber\\
&&+ \frac{\omega} {\varphi^2} (\square {\varphi}) \nabla^\mu {\varphi}
-\frac{1} {\varphi^2}  (\nabla^\mu \nabla^\nu {\varphi}) \nabla_\nu {\varphi}\nonumber\\
&&+ \frac{1} {\varphi^2} (\square {\varphi}) \nabla^\mu {\varphi} + \frac{1} {\varphi} \Big[- \frac{4 \pi}{\varphi} {T_M^\lambda}_{\lambda}{g}^{\mu \nu}
+\frac{8 \pi}{\varphi}{T_M^{\mu \nu}}~~ \nonumber\\
&&+\frac{\omega}{\varphi^2 } \nabla^{\mu} \varphi \nabla^{\nu} \varphi +\frac{1}{2 \varphi}  \square {\varphi} {g}^{\mu \nu}
+ \frac{V}{2 \varphi}{g}^{\mu \nu} + \frac{1}{\varphi} \nabla^{\mu}  \nabla^{\nu} \varphi\Big]\nabla_\nu \varphi\nonumber\\
&&+ (\frac{V}{2 \varphi^2}
- \frac{1}{2 \varphi} \frac{d V}{d \varphi}) \nabla^\mu {\varphi}=0.\label{eq25}
\end{eqnarray}
After performing some tensor calculations we find that
\begin{eqnarray}
\frac{1} {2 \varphi^2}\Big[-8 \pi {T_M^\lambda}_{\lambda} +(2 \omega +3) \square {\varphi} +2V - {\varphi}  \frac{d V}{d \varphi}\Big]
\nabla^\mu {\varphi}=0.~~\label{eq26}
\end{eqnarray}
Clearly, from the above equation we have
\begin{eqnarray}
-8 \pi {T_M^\lambda}_{\lambda} +(2 \omega +3) \square {\varphi} +2V - {\varphi}  \frac{d V}{d \varphi}=0.\label{eq27}
\end{eqnarray}
From solving Eq. (\ref{eq27}) for $\square {\varphi}$, we get the modified Klein-Gordon equation, Eq. (\ref{eq11}).
In this way, we reached the desired result.

\section{\label{V} Conclusion}

In this paper, we proved that the modified Klein-Gordon equation, Eq.  (\ref{eq11}), in the BD theory of
gravity is not an independent equation of the BD field equation, Eq. (\ref{eq8}),
but it is a result of the establishment of the BD field equation (\ref{eq8}).
Therefore, when we want to introduce the BD theory, it is enough to consider only the BD field equation, Eq. (\ref{eq8})
as the basic equation of this theory. For this reason,
if we consider Eqs.  (\ref{eq1}), (\ref{eq2}) and (\ref{eq8}) without the modified Klein-Gordon equation, Eq. (\ref{eq11})
as a system of equations in the ``problem of the FLRW cosmological model'' to determine the four unknowns
$a(t)$, $\varphi(t)$, $\rho(t)$ and $p(t)$, we have not made any mistake. Our reason for doing this is that
the modified Klein-Gordon equation, Eq. (\ref{eq11}) in the BD theory is not a fundamental equation and
only Eq. (\ref{eq8}) forms the fundamental equation of the theory.
The BD field equation, Eq. (\ref{eq8}),
provides us with two independent equations that are not enough to determine the four unknowns.
Hence, we need two other equations that together with Eq. (\ref{eq8})
form a system of four independent equations with four unknowns.
As seen in this paper, we have chosen the equation of state, Eq. (\ref{eq1})
and the power law, Eq. (\ref{eq2}), which are compatible with the physics of the problem,
as the desired equations. It is worth noting that although Eq. (\ref{eq11})
is not one of the basics equations of the BD theory, it can be added to the system of algebraic-differential Eqs. (\ref{eq1}), (\ref{eq2}) and (\ref{eq8}),
like $\nabla_\nu G^{\mu \nu}=0$ (Bianchi's identity) and $\nabla_\nu T^{\mu \nu}=0$ (conservation of the energy-momentum tensor of the matter field) without creating an obstacle in solving the FLRW cosmological model.


\end{document}